\documentstyle[prd,aps]{revtex}
\begin{document}
\title{Multiplicity difference correlators under first-order
QGP phase transition}
\author{C.B. Yang$^{1,2}$ and X. Cai$^1$} 
\address{$^1$ Institute of 
Particle Physics, Hua-Zhong Normal University, 
Wuhan 430079, P.R. China\\
$^2$ Theory Division, RMKI, KFKI,
Budapest 114., Pf. 49, H-1525 Hungary}
\date{\today}
\maketitle

\begin{abstract}
The multiplicity difference correlators between two
well-separated bins in high-energy heavy-ion collisions are studied as a
means to detect evidence of a first-order quark-hadron phase transition. 
Analytical expressions for the scaled
factorial moments of multiplicity difference distribution are obtained 
for small bin size $\delta$  with mean multiplicity in the bin
${\overline s}\le 1.0$ within
Ginzburg-Landau description. The scaling behaviors between the moments 
are still valid, though they behave completely different from the
so-called intermittency patterns. A universal exponent $\gamma=1.4066$ is given 
to describe the dynamical fluctuations in the phase transition in small
$\delta$ limit.
\end{abstract}

\pacs{{\bf PACS} number(s): 13.85.Hd, 05.70.Fh, 12.38.Mh}

The theoretical study of fluctuations in quark-hadron phase transition has been
suggested for several years$^{[1]}$. The first motivation came from the
quantitatively different results in Monte Carlo simulations $^{[2]}$
on intermittency for $pp$ collisions without phase transition
from those theoretical predictions with the onset of phase transition$^{[1]}$.
Since then, multiplicity fluctuations are studied with phase transitions
of second-order$^{[3]}$  and first-order$^{[4-6]}$ within Ginzburg-Landau
model, and it is regarded as a possible means to reveal some features for the
phase transition. Most of these works give the violation of the intermittency
patterns but show remarkable scaling behaviors between $F_q$ and $F_2$, and
there seems to exist a universal exponent $\nu$ $^{[1,3,6]}$. It is suggested
that the exponent $\nu$ can be used as a useful diagnostic tool to detect the
formation of QGP. In [4, 5] $\ln F_q$ are studied analytically for first-order phase 
transition and are expanded as power series of $\delta^{1/3}$, and it is shown
that the set of experimentally fitted coefficients for $\delta^{1/3}$ term can
be used as a criterion for the 
onset and the order of the phase transition. 

It is known for a long time that the investigation of multiplicity fluctuations
is very different in heavy-ion collisions, though the power-law dependence
of $\ln F_q$ on $\delta$, $F_q\propto \delta^{-\varphi_q}$, has been found
ubiquitous in hadronic and leptonic processes$^{[7]}$. The main differences
between heavy-ion physics and hadronic \& leptonic ones
on multiplicity fluctuations were noticed earlier in Ref. [8]. In Ref. [9, 10] 
an alternative way was proposed to study the fluctuations by means of
factorial moments of the multiplicity difference (FMMD) between two 
well-separated bins. 
This alternative is a hybrid of the usual factorial 
correlators $^{[11]}$ and wavelets $^{[12]}$ because $W_{jk}$ in Haar wavelet 
analysis is just the difference of multiplicities in two nearest bins. 
When discussing the multiplicity difference, the two bins are not necessary
the nearest ones. Instead, it will be assumed 
in present study that the two bins are well-separated.
Let the two bins, each of size $\delta^2$ and separated by $\Delta$, have 
multiplicities $n_1$ and $n_2$, and define their multiplicity difference 
$m=|n_1-n_2|$. Let $Q_m$ be the distribution of multiplicity difference, 
which may be dependent on $\Delta, \delta$ and details of the process. 
Scaled FMMD are defined as
\begin{equation}
{\cal F}_q=f_q/f_1^q, \mbox{\hspace{0.8cm}}
f_q=\sum_m m(m-1)\cdots (m-q+1) Q_m\ ,
\end{equation}

\noindent Moments defined above are similar to but not the same as the 
Bialas-Peschanski correlators $^{[11]}$ $F_{q_1q_2}$, because of 
(1) ${\cal F}_q$ are moments of the multiplicity difference
between two bins and (2) ${\cal F}_q$ may depend on both $\Delta$
and $\delta$ while $F_{q_1q_2}$ depends only on $\Delta$. 

In Ref. [9], ${\cal F}_q$ are numerically studied within
Ginzburg-Landau model. The scaling behaviors between ${\cal F}_q$ and 
${\cal F}_2$, ${\cal F}_q\propto {\cal F}_2^{\beta_q}$, are shown with
$\beta_q=(q-1)^\gamma$ and a universal exponent $\gamma$=1.099.
In Ref. [10], ${\cal F}_q$ are analytically studied for very small
$\delta$ within the same model for second-order phase transition.
The dynamical components of the moments are introduced and are shown to
have similar scaling behaviors. The universal exponent $\gamma$=1.3424 given
in [10] for dynamical fluctuations is shown to be different from that in [9]
but close to $\nu$ given in the study of usual factorial moments$^{[3,6]}$. 
The closeness of $\gamma$=1.3424 to $\nu\simeq 1.30$ is reasonable because
they both describe dynamical fluctuations in phase transition. The little
difference comes from the different $x$ regions concerned: $\nu$ corresponds
to $x$ region around $-\ln x\sim 1$ but $\gamma$ to $x\to 0$, as discussed 
in [10]. The most important feature about this $\gamma$ is that it is
completely determined by the general 
features of the model and independent of the parameters for the model.

In this Letter, scaled FMMD ${\cal F}_q$ in first-order phase transition 
will be studied analytically for very small bin size
$\delta$. For simplicity, the discussion is limited to two identical
small bins under the condition that the mean multiplicity in each bin is less
than or equal to 1.0.

As a base and starting point, discuss the trivial and simplest case in which
there is no correlations between the two bins and within each bin.
Let the mean multiplicity in 
each bin is $s$. Because no dynamical reason is assumed, 
the multiplicity distribution for each bin is a Poisson one
\begin{equation}
P_{n}(s)={s^{n}\over n!} \exp(-s)\ \ .
\end{equation}

\noindent From this distribution, one can deduce the multiplicity difference
distribution as
\begin{equation}
P_m(s)=I_m(2s)e^{-2s}\ (2-\delta_{m0})\ ,
\end{equation}

\noindent where $I_m(z)$ is the modified Bessel function of order $m$,
\begin{eqnarray*}
I_m(z)=\sum_{k=0}^{\infty} {(z/2)^{2k+m}\over k!(k+m)!}\ .
\end{eqnarray*}

\noindent FMMD for pure statistical fluctuations are
\begin{equation}
f_q^{\rm (stat)}=\sum_{m\ge q} m(m-1)\cdots (m-q+1) P_m(s)\ .
\end{equation}

To complete the summation in last equation, one can introduce a generating
function$^{[9]}$
\begin{equation}
G(x,s)=2e^{-2s}\sum_{m=0}^\infty x^m I_m(2s)\ ,\mbox{\hskip 0.5cm}
G_q(x, s)={d^q G(x,s)\over dx^q} \ .
\end{equation}

\noindent With this function, $f_q^{\rm (stat)}$ can be rewritten as
\begin{equation}
f_q^{\rm (stat)}=G_q(1,s)\equiv G_q(s)\ .
\end{equation}

\noindent Direct algebra shows that
\begin{equation}
G(x,s)=2e^{(x-2)s}\left[a_0^0+\sum_{i=1}^\infty a_i^0 {d^i\over dx^i}
{1-\exp(-xs) \over x}\right]
\end{equation}
 
\noindent with $a_i^0=(-1)^i s^{2i}/(i!)^2$ for $i=0, 1, \cdots$, and that
\begin{equation}
G_q(s)=2e^{-s}\left[a_0^q+\sum_{i=1}^\infty a_i^q\sum_{j=i}^\infty
{(-1)^j s^{j+1}\over (j+1)(j-i)!}\right]\ ,
\end{equation}

\noindent where $a_i^q$ can be calculated by recurrence relation from
$a_i^0$, $a_0^q=sa_0^{q-1}$, $a_1^q=sa_1^{q-1}$, $a_i^q=sa_i^{q-1}+
a_{i-1}^{q-1}\ , (i\ge 2)$. 

In this Letter, we are interested only in small bin analysis, for which
the mean multiplicity is less than 1. Then $G_q(s)$ can be approximately
written as
\begin{equation}
G_q(s)=2{\rm e}^{-s} s^q\left[1+{s^4\over (q+1)(q+2)}-
{2s^5\over (q+1)(q+2)(q+3)}+{3s^6\over (q+1)(q+2)(q+3)(q+4)}\right]\ ,
\end{equation}

\noindent and all other terms can be omitted. Direct estimation shows
that the error caused by this approximation is less than 1\% for $s\le 1.0$. 

For cases with dynamical fluctuations due to phase transition, the distribution of
multiplicity difference is$^{[9]}$
\begin{equation}
Q_m(\delta, \tau)=Z^{-1}\int {\cal D}\phi P_m(\delta^2\tau\mid\phi\mid^2)e^{-F[\phi]}\ ,
\end{equation}

\noindent where $\tau$ is an indication of lifetime of the whole parton system,
${\cal D}\phi=\pi d\mid\phi\mid^2, Z=\int {\cal D}\phi e^{-F[\phi]}$ and
the free energy $F[\phi]=\int_{\delta^2} dz\left[a\mid\phi\mid^2+b\mid\phi\mid^4
+c\mid \phi\mid^6\right]$ for first-order phase transition.

Substituting $Q_m(\delta, \tau)$ into Eq. (1), one gets
\begin{equation}
f_q=\left.\int_0^\infty dy\, G_q(\tau ux^2y)\,{\rm e}^{-y^3+ux^2y+xy^2}\right/
\int_0^{\infty} dy\, {\rm e}^{-y^3+ux^2y+xy^2}
\end{equation}

\noindent with $x=-b(\delta/c)^{2/3}$ related with the bin width $\delta$,
$u=\mid a\mid c/b^2$. As discussed in Ref. [4-6], $b$ is negative for first-order
phase transition. So $x$ and $u$ are both positive in present discussions.
For the convenience of analytical calculations, define$^{[4]}$
\begin{equation}
H_q(u,v)=\int_0^\infty y^q\exp(-y^3+uv^2y+vy^2) 
\end{equation}

\noindent which satisfies recurrence relations
\begin{eqnarray*}
H_2(u,v)&=&{1\over 3}+{1\over 3}\left(uv^2H_0(u,v)+2vH_1(u,v)\right)\ ,\\
H_{q+3}(u,v)&=&{1\over 3}\left[(q+1)H_q(u,v)+uv^2H_{q+1}(u,v)+2vH_{q+2}(u,v)\right]\ .
\end{eqnarray*}

 \noindent With $H_q(u,v)$, $f_q$ can be expressed as
\begin{equation}
f_q={2\over H_0(u,x)}\left[H_q+{(\tau ux^2)^4H_{q+4}\over (q+1)(q+2)}
-{2(\tau ux^2)^5H_{q+5}\over (q+1)(q+2)(q+3)}+{
3(\tau ux^2)^6H_{q+6}\over (q+1)(q+2)(q+3)(q+4)} \right] \ .
\end{equation}

\noindent all $H_q$ in the bracket in last equation should be read as 
$H_q(-(\tau-1)u, x)$. 

The scaled FMMD ${\cal F}_q$ obtained contain contributions from statistical
fluctuations, contrary to the usual scaled factorial ones. One can see this
clearly if one notices the fact that ${\cal F}_q$ are not equal to 1
for the case with pure statistical fluctuations. To eliminate the
pure fluctuations from the moments,  
one can define the dynamical scaled FMMD as$^{[10]}$
\begin{equation}
{\cal F}_q^{(\rm dyn)}={{\cal F}_q\over {\cal F}_q^{\rm (stat)}}\ .
\end{equation}

\noindent To make the definition sense, one should ensure that the mean
multiplicity is the same for all the calculation of the moments concerned. 
In Ginzburg-Landau model, the mean multiplicity is ${\overline s}=
\tau ux^2\,H_1(u, x)/H_0(u,x)$. Then deviations of ${\cal F}_q^{(\rm dyn)}$
from one should indicate the existence of dynamical fluctuations.
Different from the case for second-order phase transition, ${\overline s}
\propto x^2$ in small $x$ region for first-order phase transition, thus the
pure statistical fluctuations have no contribution to the slopes of $\ln
{\cal F}_q$ in the $x$ region we are now interested in. Thus, we will only
discuss $\ln {\cal F}_q$ in present discussion.

The dependences of $\ln {\cal F}_q$ on $-\ln x$ from 2.0 to 4.0 are shown
in Fig. 1 for $\tau$=10.0, $u=1.0$ and 5.0. The $-\ln x$ range
is chosen from the requirement that ${\overline s}$ is much less
than 1.0 for $\tau=10.0$ and $u=5.0$. 
From this figure, one can see clearly that $\ln F_q$  
depend strongly on parameters chosen, and $\ln {\cal F}_q$ decrease 
quickly with the increase of $-\ln x$ in the $x$ range chosen. 
For sufficiently large $-\ln x$, $\ln {\cal F}_q$
satuate to values independent of $u$. In fact, the satuation values are
independent of any parameter in the model.

Scaling behaviors between ${\cal F}_q$ are shown in Fig. 2
for the same choices of parameters as in Fig. 1.
One can see that the scaling behavior, ${\cal F}_q\propto {\cal F}_2^{\beta_q}$,
is valid for either set of the parameters, and the slopes show weak dependence
Because of the linearity of curves
in Fig. 2 for ${\cal F}_q$, $\beta_q$ can be determined
accurately. One can guess that for small $x$, $\beta_q$ may be independent
of all parameters in the model, and it is indeed the case. For very small $x$,
one can get, using $H_q(u,v)\simeq {1\over 3}\left[ \Gamma({q+1\over 3})+v
\Gamma(1+{q\over 3})\right]$
\begin{equation}
\beta_q^{\rm (dyn)}={(q-1)a_{0,1}/a_{0,0}+a_{q,1}/a_{q,0}-q a_{1,1}/a_{1,0}\over
a_{0,1}/a_{0,0}+a_{2,1}/a_{2,0}-2 a_{1,1}/a_{1,0}}\ ,
\end{equation}

\noindent
with $a_{q,0}=\Gamma({q+1\over 3})$ and $a_{q,1}=\Gamma(1+{q\over 3})$.
$\beta_q$ as function of $\ln (q-1)$ is shown in Fig. 3. A perfect
scaling behavior is shown 
\begin{equation}
\beta_q=(q-1)^\gamma
\end{equation}

\noindent with $\gamma$=1.4066. This exponent corresponds to the limit $x\to 0$.
In real experimental analysis, one fits curves for $\ln {\cal F}_q$
and $\ln {\cal F}_2$ at finite $x$, so that one should get an
exponent with a little difference from the one given here. In fact,
the experimentally obtained exponent should be less than 1.4066.

It should be pointed out that the exponent can be obtained simply from the leading
term in $G_q$ by dropping off all non-leading terms because they are related to
higher orders of $x$ and have no contribution to $\gamma$ which
is connected with properties of the moments in the limit $x\to 0$. 
The exponent given here is different from that in [10] for second-order phase 
transition. The main reason is due to the introduction of $\mid\phi\mid^6$
term in the free-energy. One can see that $\mid\phi\mid^6$ term is necessary
for the study of first-order phase transition. In the case of second-order phae 
transition, the same term may also play a role. Only when $-x\gg 1$ can the $-y^3$ 
term be neglected in Eq. (11), and one returns to the Ginzburg-Landau description
of the phase transition studied in [3]. So the discussions in this Letter
are different from those in former studies. In fact, if $\mid\phi\mid^6$ term
plays an important role, $\gamma$ should be close to 1.4066 for both
first-order and second-order phase transitions in small $\delta$ analyses.
With the increase of the importance of $\mid\phi\mid^6$ term, the
exponent $\gamma$ for second-order phase transition can undergo an
 increase from 1.3424 in [10] to 1.4066 given here.
But if there is no phase transition, the exponent $\gamma$ should be much less
than 1.3424. Thus if an exponent $\gamma$ for the dynamical fluctuations is
found near to 1.4, then the occurrence of quark-hadron phase transition can
be pronounced. 

In summary, scaled FMMD are studied analytically within Ginzburg-Landau model
in a kinetical region with mean multiplicity in single bin less than 1.0 for
first-order quark-hadron phase transition. The dynamical fluctuations in FMMD
are extracted, which give the same physical contents as the usual scaled factorial
moments. Scaling behaviors between scaled FMMD are shown, and a truly universal
exponent $\gamma=1.4066$ is given.

This work was supported in part by the NNSF, the SECF and Hubei NSF in China.
One of the authors (C.B. Yang) is grateful for fruitful discussions with
Prof. R.C. Hwa.

\centerline{{\Large Figure Captions}}
\begin{description}
\item
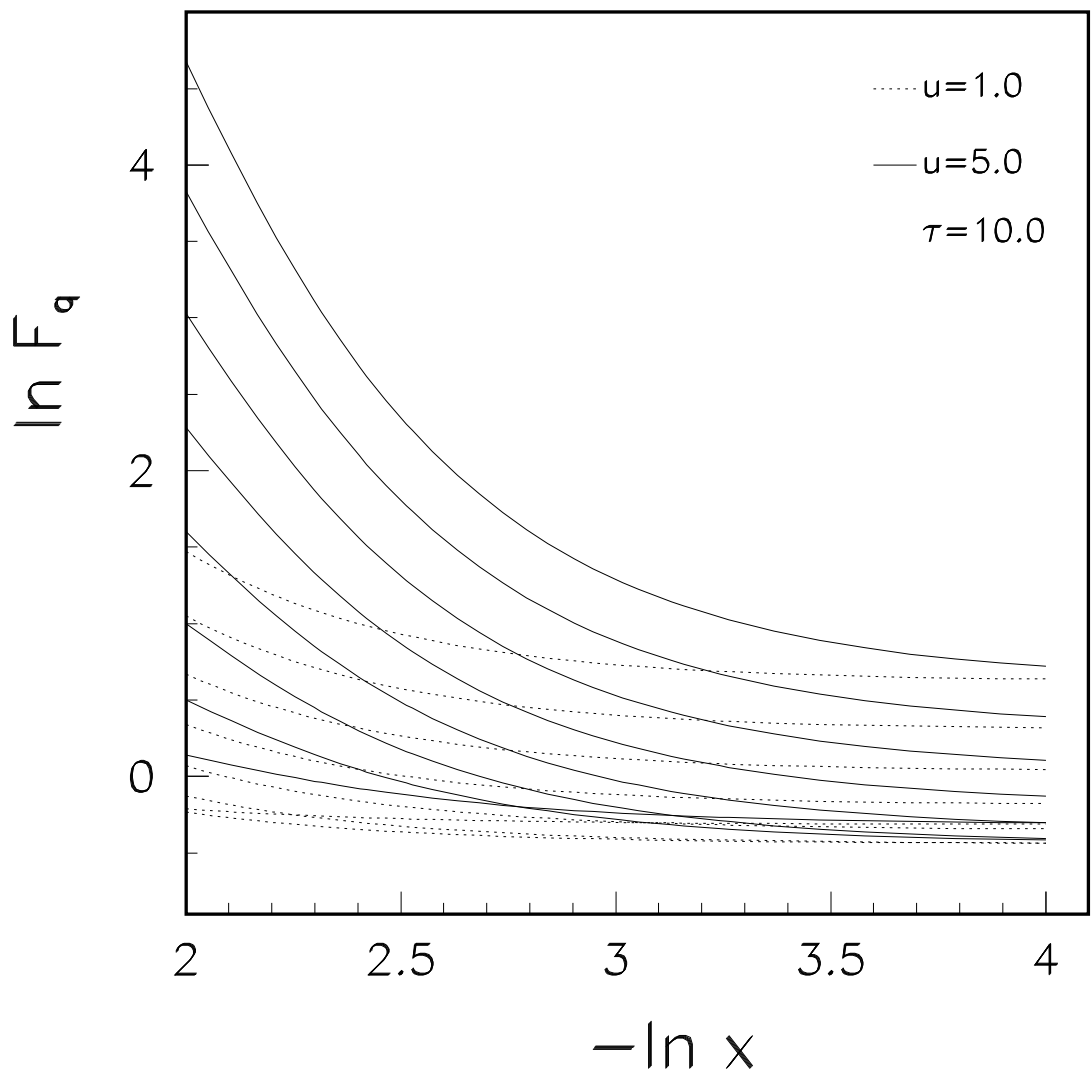
{\bf Fig. 1} Dependences of $\ln {\cal F}_q$
on $-\ln x$ for $\tau=10.0$, $u=1.0$ and $u=5.0$.
From lower to upper are curves for $q$=2,3,4,5,6,7,8,9, respectively.
\item
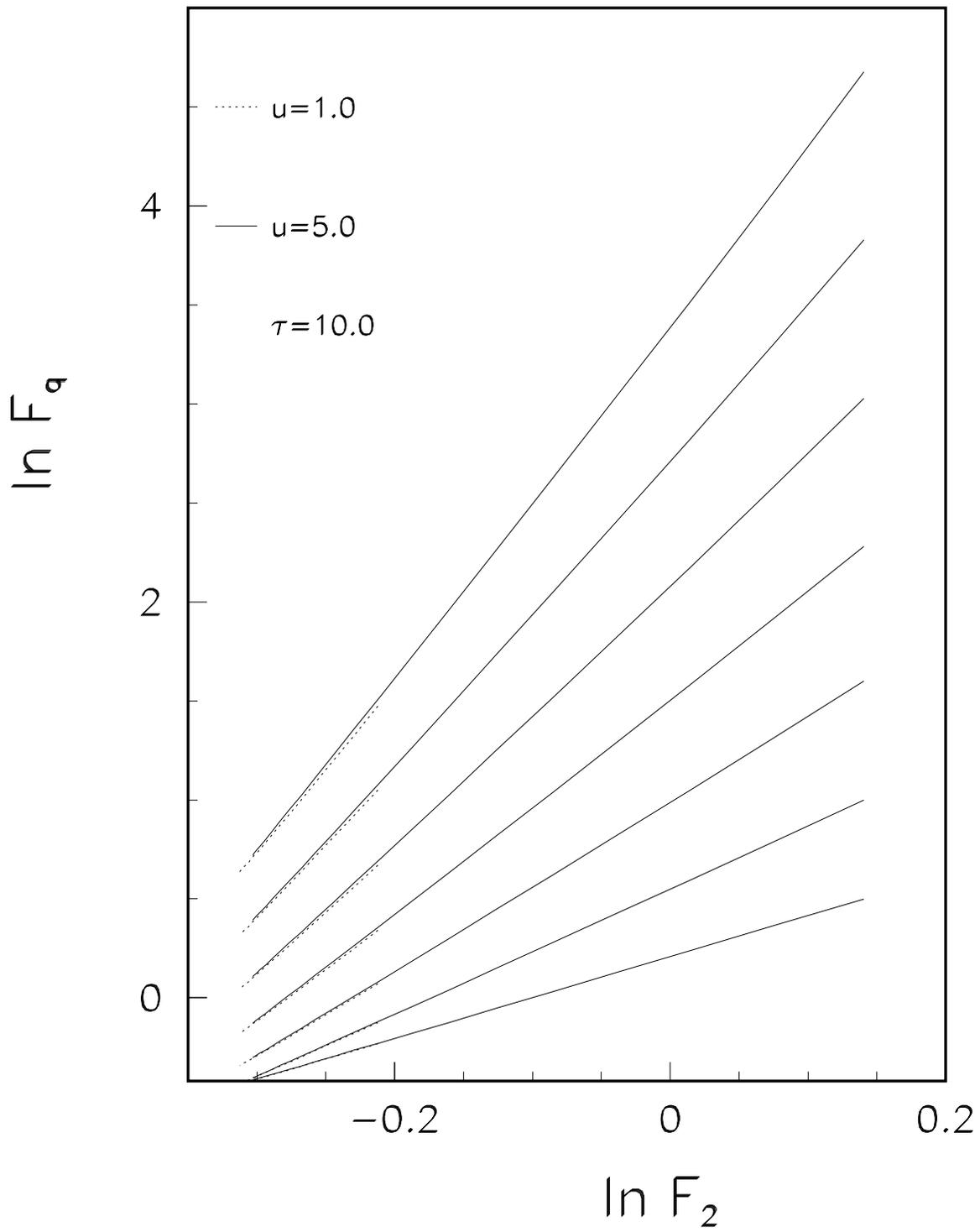
{\bf Fig. 2} Scaling behaviors of $\ln {\cal F}_q$ vs
$\ln {\cal F}_2$ for the same choices of parameters
as in Fig.1. 
From lower to upper are curves for $q$=3,4,5,6,7,8,9, respectively.
\item
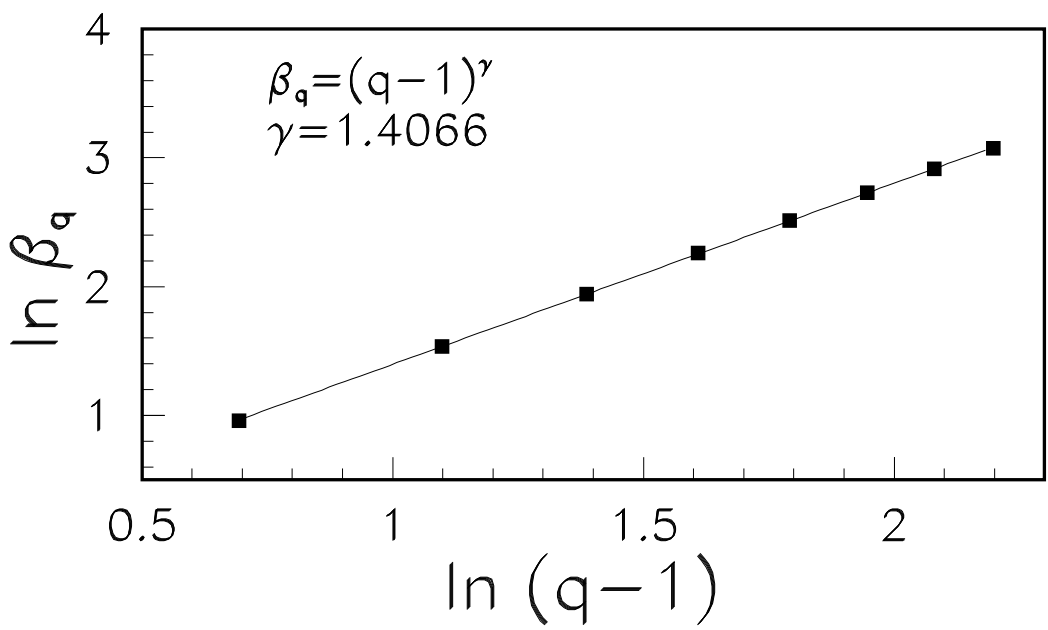
{\bf Fig. 3} Scaling behavior of $\ln \beta_q$ vs $\ln (q-1)$.
\end{description}
\end{document}